\documentclass[3p,times,twocolumn]{elsarticle}

\usepackage{ecrc}

\volume{00}

\firstpage{1}

\journalname{Nuclear Physics B Proceedings Supplement}

\runauth{S.~Stefanik and D.~Nosek}

\jid{nuphbp}

\jnltitlelogo{Nuclear Physics B Proceedings Supplement}

\usepackage{amssymb}

\newcommand{\on}{N_{\mathrm{on}}}
\newcommand{\off}{N_{\mathrm{off}}}
\newcommand{\Bi}{S_{\mathrm{Bi}}}
\newcommand{\LM}{S_{\mathrm{LM}}}
\newcommand{\pks}{PKS~2005--489}
\newcommand{\es}{1ES~0229+200}

\begin{document}

\begin{frontmatter}

\title{Variability of VHE $\gamma$--ray sources}
\author[ipnp]{Stanislav Stefanik\corref{cor1}}
\cortext[cor1]{Corresponding author.}
\ead{stefanik@ipnp.troja.mff.cuni.cz}
\author[ipnp]{Dalibor Nosek}
\ead{nosek@ipnp.troja.mff.cuni.cz}
\address[ipnp]{Institute of Particle and Nuclear Physics, Faculty of Mathematics and Physics, Charles University \\ 
V~Holesovickach~2, 180 00 Prague 8, Czech Republic}


\begin{abstract}
We study changes in the $\gamma$--ray intensity at very high energies observed from selected active galactic nuclei.
Publicly available data collected by Cherenkov telescopes were examined by means of a simple method utilizing solely the number of source and background events.
Our results point to some degree of time variability in signal observed from the investigated sources.
Several measurements were found to be excessive or deficient in the number of source events when compared to the source intensity deduced from other observations.
\end{abstract}


\end{frontmatter}



\section{Introduction}
\label{intro}
Precise knowledge of variability in $\gamma$--ray fluxes observed from various emitters is considered essential for probing their intrinsic properties.
Among the most prominent sources of such transient $\gamma$--ray emission are active galactic nuclei (AGN).
Measurements of variability time scales of their activities can be useful, for example, for putting constraints on size and location of the $\gamma$--ray production region~\citep{Sol:2013}.
\par 
Spectral energy distributions (SED) of AGN exhibit two distinctive peaks and they are commonly described by the synchrotron self--Compton model (SSC)~\citep{Katarzynski:2001}.
This scenario assumes that a population of relativistic electrons present in the AGN jets gives rise to the synchrotron photons at X--ray wavelengths in the vicinity of the first SED peak.
High energy radiation of the second spectral bump is believed to be a result of inverse Compton scattering of lower energy photons by the same ensemble of electrons that is responsible for the synchrotron emission.
Information on temporal changes of different SED components of AGN and their mutual relations can be beneficial for predictions on the mechanisms of the particle acceleration and photon emission~\citep{Sol:2013}.
\par 
In particular, analysis of data gathered during observations of the \pks~blazar provided no conclusive proof of the flux variability at very high energies (VHE) in the GeV--TeV $\gamma$--ray band between years 2004 and 2005~\citep{Acero:2010}.
This is contrary to the findings at other wavelengths which indicate that the X--ray flux  increased significantly during the two consecutive years~\citep{Acero:2010}.
If the VHE radiation is to be related to the lower energy contributions to the SED through the SSC mechanism, a rise in the X--ray activity should be accompanied by corresponding increase of the flux in the TeV band~\citep{Katarzynski:2001}.
The lack of the $\gamma$--ray flux variability might be explained by an additional jet component of electrons contributing to the VHE emission~\citep{Acero:2010}.
Such component, emerging only in the hard X--ray energy range, should be separated from the production region of the observed synchrotron emission and thus not interact with these photons in order to preserve the measured VHE flux~\citep{Acero:2010}.
\par 
Besides studying the origin of $\gamma$--ray emission, possible time variations in the source intensities of hard--spectrum blazars located at large redshifts can have also implications for cosmological observations.
Constraints can be put on the strength of the intergalactic magnetic field (IGMF) from measurements of time delay between arrivals of photons in different energy bands as long as the emission of the parent very high energetic $\gamma$--rays is steady \citep{Dermer:2011}.
The condition of the constant flux thus calls for reliable investigations of multiwavelength variability of AGN.
Distant blazar \es~has been for a long time considered to be a good candidate for IGMF studies~\citep{Dermer:2011, Neronov:2010}.
This belief has been based on apparent steadiness of the VHE $\gamma$--ray flux observed during 2005--2006~\citep{Aharonian:2007}.
However, analysis of more recent data casts doubt upon this assumption with the recommendation not to include this object in the IGMF research relying on a constant flux or at least account for systematic uncertainties arising from the variability of its flux~\citep{Aliu:2014}.
The case of \es~is thus another example of strong need for precise measurements and reliable statistical methods for variability studies.
\par
Future improvement in the experimental techniques in $\gamma$--astronomy will provide us with vast amount of data on various transient phenomena with better temporal resolution.
In particular, one of the goals of the next generation of imaging atmospheric telescopes, the Cherenkov Telescope Array~\citep{Acharya:2013}, will be extensive studies of the AGN populations~\citep{Sol:2013}.
Time variability of intrinsic activities of not only these $\gamma$--ray emitters will be a crucial question in most analyses, for which our modification~\citep{Nosek:2013} of the no--source on--off method~\citep{Li:1983} could be useful.
\par
The modified on--off technique attempts to determine a level of significance for an excess or deficit of counts in individual measurements when compared to the reference source intensity previously ascertained from other observations~\citep{Nosek:2013}.
In the on--off method there is no demand for the calculation of the flux or other quantities.
It works only with the numbers of events detected in the on--source and reference off--source region provided that their exposures are known.
The method is equally suitable for any observations regardless of the experimental technique, thus making the comparison of data detected by different instruments possible.
\par
The modified on--off method is briefly described in Section~\ref{sec:method}.
It allows to check for intensity changes in the ranges of times, energies or zenith angles, for example, as long as reasonable estimates for the source intensity exist.
Section~\ref{sec:analysis} deals with the analysis of the data gathered by experiments using imaging atmospheric Cherenkov telescopes along with the previous findings on the considered sources.
We use the on--off technique to examine whether two selected AGN, \pks~and \es, exhibit any temporal changes in the numbers of $\gamma$--ray events detected from their directions.


\section{Method}
\label{sec:method}

The Li--Ma method is widely employed in VHE $\gamma$--ray astronomy for determining a level of significance of a photon excess above background when validating the source presence in a given region \citep{Li:1983}.
In this method, one assumes the test of the null hypothesis stating that there is no source present in the investigated on--source region.
The on--source region encompasses the potential $\gamma$--ray emitter whereas the off--source region is considered to be free of point sources and thus suitable for the background estimation.
In order to account for different extent (e.g.~temporal or spatial) and unequal observing conditions of both regions, an on--off parameter $\alpha$, the ratio of the on-- and off--source exposures, is needed. 
\par 
A straightforward modification of this technique allows one to estimate the significance of an excess or deficit of the number of events when compared to the known source activity \citep{Nosek:2013}.
The modified on--off method assumes that the $\gamma$--ray emitter has already been positively identified in the potential hotspot.
A source parameter $\beta > 0$ is introduced to characterize its intensity.
Then the statement of the null hypothesis is that the source attains intensity of previously chosen value~$\beta$, i.e.~$\on = \alpha \beta \off$~\citep{Nosek:2013}, where $\on$ and $\off$ are the numbers of events detected in the on-- and off--source regions, respectively.
\par 
A level of significance for the rejection of the no--source assumption is given in terms of either binomial or Li--Ma statistics \citep{Li:1983}.
A modification of the original significance formulae (Eqs.~9 and 17 in \citep{Li:1983}) for the assumption of the constant source intensity leads to a couple of similar equations, the only difference being the transformation $\alpha \rightarrow \alpha \beta$ \citep{Nosek:2013}, i.e.

\begin{equation}
\label{eq:bin}
\Bi = \frac{\on - \alpha \beta \off}{\sqrt{\alpha \beta \left( \on + \off \right) }},
\end{equation}

\begin{equation}
\label{eq:lima}
\LM = s\sqrt{2}\left\lbrace \on \ln{ X_{1} } + \off \ln{ X_{2} } \right\rbrace^{\frac{1}{2}},
\end{equation}

 

\noindent
for the binomial and Li--Ma statistics, respectively.
Here, the logarithmic arguments are $X_{1} = \frac{1+\alpha\beta}{\alpha\beta} \frac{\on}{\on + \off}$ and $X_{2} = (1+\alpha\beta) \frac{\off}{\on + \off} $.
The $s$--term in Eq.~\ref{eq:lima} ($s = \pm 1$) accounts merely for the sign of the whole expression, depending whether an excess ($\Bi,~\LM>0$) or deficit ($\Bi,~\LM<0$) of events is observed.
\par 
Taking the source parameter $\beta$ being equal to unity one recovers the original no--source assumption.
Alternatively, the inequality \mbox{$\beta > 1$} expresses that an excess of counts above the source intensity will be tested while \mbox{$0 < \beta < 1$} implies the test of their deficit.
\par
It can be shown that the binomial and Li--Ma significances written in Eqs.~\ref{eq:bin} and \ref{eq:lima} are to be considered asymptotically as drawn from the normal distribution with zero mean and unit variance \citep{Nosek:2013}.
Therefore, any inconsistency between the sample distributions of $\Bi$ and $\LM$ and the reference standard Gaussian distribution should be regarded as a sign of change in the tested $\gamma$--ray intensity.
It is worth noting, that the Li--Ma statistic can be alternatively exploited to derive confidence intervals for the source parameter $\beta$ at a given level of significance.
Sequence of such confidence intervals carries information about temporal evolution of the source $\gamma$--ray activity.


\section{Data analysis}
\label{sec:analysis}

We used published data obtained during observations of two VHE $\gamma$--ray emitters by imaging atmospheric Cherenkov telescopes.
The only necessities for the modified on--off method are the numbers of detected on-- and off--source events, $\on$ and $\off$, and the on--off parameter $\alpha$.
These quantities are the direct outcome of the experiment and its particular setup. 
On contrary, the source parameter $\beta$ is set at our discretion.
We used estimates of the source parameter derived as average values of the ratio of observed and expected on--source events over individual time intervals, i.e.~$\beta = \langle \on / \alpha \off \rangle$.
Using a particular value of the source parameter, the investigated data were tested for the assumption of given intensity.
\par 
In the following, our results are visualized in quantile--quantile (QQ) plots, see Figs.~\ref{fig:PKS2005_QQ} and \ref{fig:1ES0229_QQ}.
In these plots, both binomial and Li--Ma significances retrieved from data are arranged in ascending order and then paired with the quantiles of the reference Gaussian distribution with zero mean and unit variance.
In order to cover evenly the unit interval, the quantiles of the standardized normal distribution were chosen to be \mbox{$k / (m+1)$} where $m$ denotes the number of observed events and $k = 1 \dots m$.
In QQ--plots, the observational time sequence is indicated as increasing sizes of markers.
\par 
In the case of a steady $\gamma$--ray intensity, the values of binomial and Li--Ma statistics lie on the diagonal of the first quadrant represented by a dashed line in QQ--plots provided the source parameter was chosen correctly.
Disagreement between the observed data and the chosen source intensity manifests itself in QQ--plots as dispersion of sample significances from the reference diagonal.
Those distributions of sample significances which provide curved QQ--plots (differently skewed than the reference standard normal distribution) indicate a progressive change in the source intensity.
The non--linear relationship between the sample and reference distributions is thus regarded as a hint of the time variability of the source activity.
\par 
In individual observations, we consider a large value of the statistical significance as a signature of a considerable deviation of the source intensity from its predefined constant value.
The upward (downward) shift of the sample statistic from the reference distribution suggests that the intensity of the source was found to be above (below) the benchmark value.
\par
Due to the arbitrariness of the source intensity represented by the source parameter $\beta$ we derived also its confidence intervals at a $3\sigma$ ($\approx 99.7\%$) level of significance.
For each triplet $(\on,~\off,~\alpha)$, confidence intervals for the source parameter $\beta$ were determined numerically such that the Li--Ma significance in Eq.~\ref{eq:lima} satisfies $|\LM(\on,~\off,~\alpha;~\beta)|<3$.
Plots of 99.7\% confidence intervals, $\langle \beta_{-}, \beta_{+} \rangle$, were constructed as MJD--ordered time sequence, see Figs.~\ref{fig:PKS2005_CI} and \ref{fig:1ES0229_CI}.
In these plots, comparison of the source intensity in individual observations is possible, thus allowing one to see the progress of the $\gamma$--ray activity in time.


\subsection{PKS 2005--489}

\par
High--frequency peaked BL Lac (HBL) object \pks~($z=0.071$) was detected by the HESS~te\-le\-scope array~\citep{Acero:2010} in each year during the 2004--2007 observational period.
The source was generally found in a state of low VHE emission.
Weak changes of the integral flux above the energy threshold 400~GeV were discovered on time scales of years, months in 2006 and nights in 2005 and 2006~\citep{Acero:2010}.
It was reported that no flux variability was observed in other dark periods but the intensity changes comparable in amplitude with statistical uncertainties cannot be excluded.
The HESS~collaboration found that the integral $\gamma$--ray flux changed annually by less than $40\%$ between 2004 and 2005.
On the other hand, observations by the satellites XMM--Newton and RXTE indicate variations of the X--ray flux by a factor of $\sim16$ during the same period~\citep{Acero:2010}.
It was also found that the flux increased by approximately $40\%$ in the ultraviolet and by $20\%$ in the optical band.
\par
\begin{figure}[!t]
	\begin{center}
		\includegraphics[width=\columnwidth]{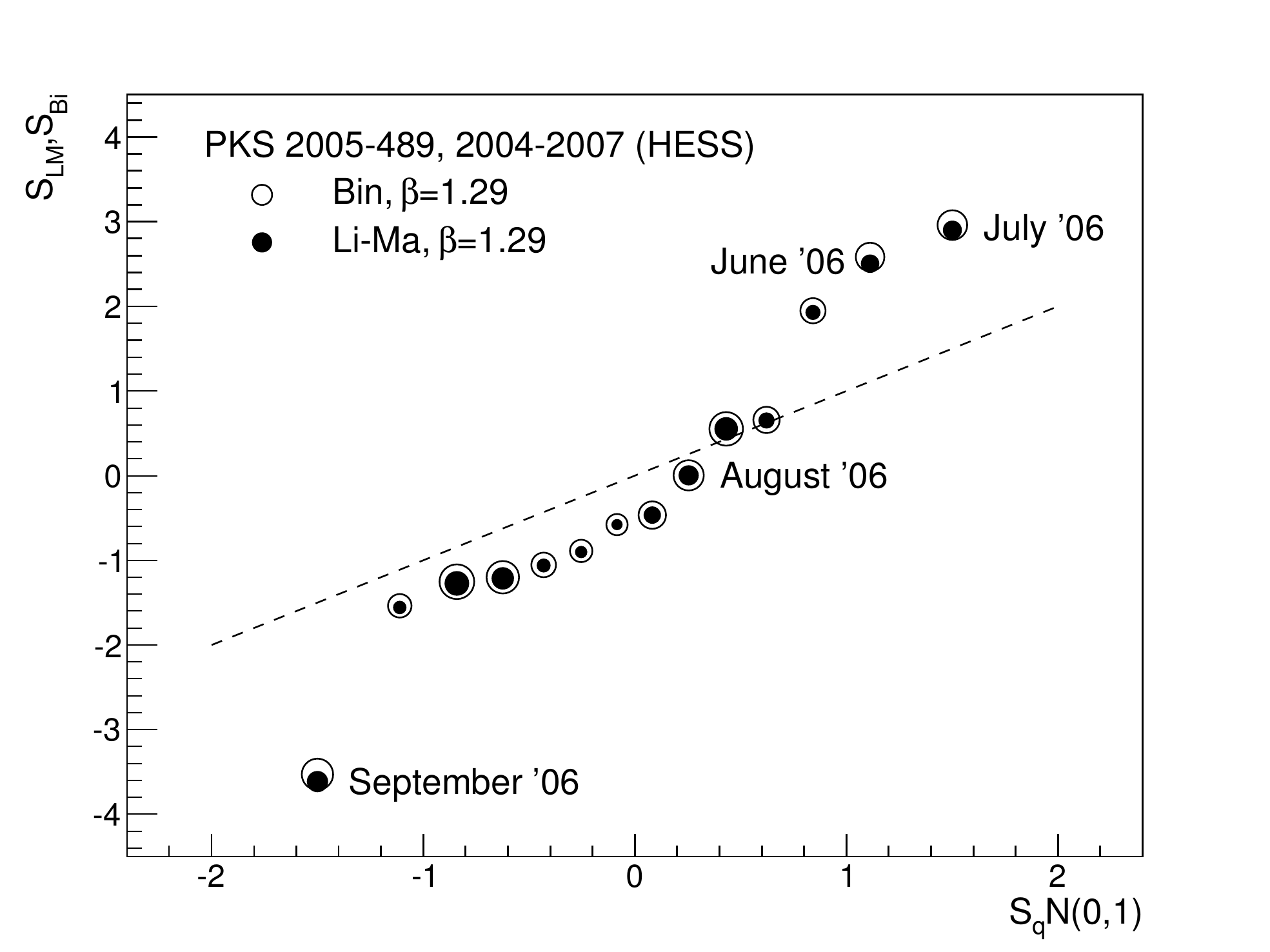}
		\caption{
			QQ--plots of asymptotic binomial (empty symbols) and Li--Ma significances (full symbols) for the $\gamma$--ray events detected from the direction of PKS~2005--489~\citep{Acero:2010}.
			Average source parameter $\beta = 1.29$ was assumed.
			Increasing sizes of markers represent the chronological sequence of observations.
			Dashed line denotes the reference diagonal of the first quadrant.
		}
		\label{fig:PKS2005_QQ}
	\end{center}
\end{figure}

 
In our analysis, we used 14 sets of data from Table~1 in Ref.~\citep{Acero:2010} arranged according to the calendar months during four years of HESS~observations.
Numbers of on-- and off--source events detected in these periods range from 93 to several thousand.
The data taken during September 2007 were not included in the analysis due to the small number of registered on--source events $\on = 11$.
We set the average source parameter $\beta = \langle \on / \alpha \off \rangle = 1.29$ as a benchmark value of the blazar $\gamma$--ray intensity.
Resultant sample significances are depicted in QQ--plots in Fig.~\ref{fig:PKS2005_QQ}.
\par 
Curvatures of QQ--plots indicate that the $\gamma$--ray intensity evolved in time during the observations. 
Dispersed distributions of the sample significances show that the intrinsic activity of the blazar is not consistent with the chosen intensity.
One pair of sample statistics, well below the reference diagonal, suggests that a deficit of $\gamma$--ray events was observed during September 2006 at a $3.2\sigma$ level of significance.
On the other hand, two measurements of June and July 2006 can be considered excessive compared to the average expectation at a $2.7\sigma$ and $3.2\sigma$ level of significance, respectively.
We conclude that the \pks~blazar exhibited variability of its emissive intensity on time scales of months during 2006.

\begin{figure}[!t]
	\begin{center}
		\includegraphics[width=\columnwidth]{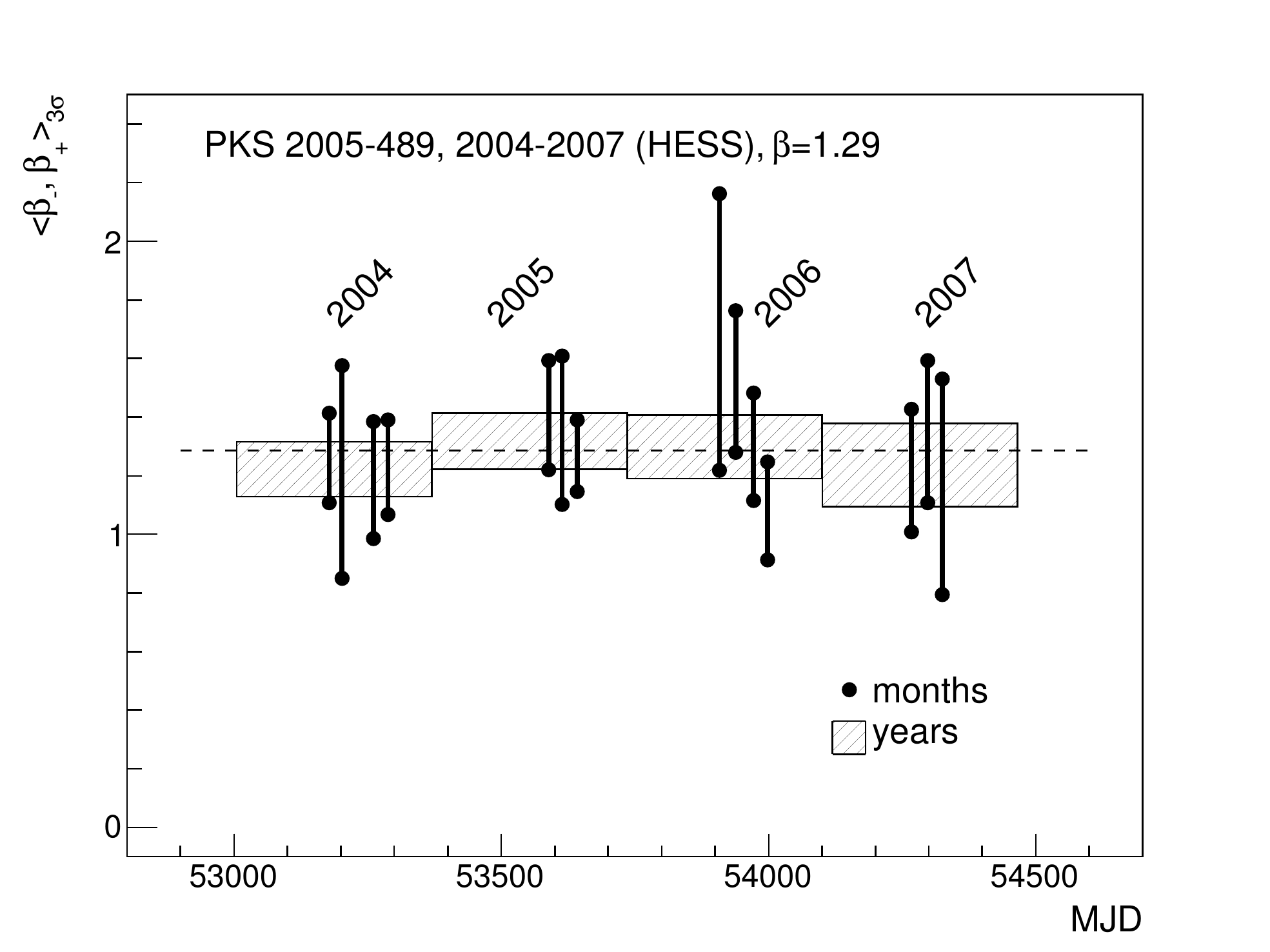}
		\caption{
			$99.7\%$ confidence intervals for the source parameter $\beta$ of \pks~\citep{Acero:2010} are shown as a function of observational time.
			Vertical lines with points indicate the span of monthly confidence intervals.
			Hatched bands represent confidence intervals in calendar years from 2004 to 2007.
			The average source parameter $\beta=1.29$ is depicted as the horizontal dashed line.
		}
		\label{fig:PKS2005_CI}
	\end{center}
\end{figure}

\par
Confidence intervals for the source parameter $\beta$ were determined for each of the 14 observational epochs, see~Fig.~\ref{fig:PKS2005_CI}.
Non--overlapping confidence intervals obtained using July and September 2006 data (MJD 53938--53940, 53995--54002) indicate consistently with corresponding points in QQ--plots in Fig.~\ref{fig:PKS2005_QQ} that a change of the source intensity happened during this period.
Observations of August 2006 (MJD 53967--53977) provide possible values of the source parameter consistent with both the preceding and subsequent measurements.
Therefore, there is a considerable hint that a gradual decrease of the source intensity occured during consecutive months in 2006, see~Figs.~\ref{fig:PKS2005_QQ} and \ref{fig:PKS2005_CI}.
The rest of measurements provide overlapping confidence intervals containing the average value of the source parameter (dashed line).
The data thus suggests that the HBL object \pks~ spent most of the time during the observational campaign in a quiescent steady state.
We evaluated also confidence intervals for the sum of counts detected in individual calendar years (hatched bands in Fig.~\ref{fig:PKS2005_CI}).
All four annual confidence intervals correspond with each other as well as with the reference source intensity.
No variations of the blazar activity can be recognized on annual time scales.
\par 
Lack of correlations between the increase of the flux at lower wavelengths (optical to X--ray) and the steady VHE emission during 2004--2005 would spoil the connection between these energy bands observed commonly in a number of blazars~\citep{Katarzynski:2001}.
In order to overcome this problem, the HESS~collaboration suggested that an additional component contributing to the X--ray part of the SED might be present in the AGN jet in a way being consistent with the SSC modelling~\citep{Acero:2010}.
We confirmed the previous HESS statement of constant $\gamma$--ray activity by analysing the sequence of confidence intervals using the modified on--off method.
We also verified that confidence intervals inconsistent with the assumption of steady source activity corresponding to the 2004 and 2005 measurements are obtained only when constructed at most at a $1.5\sigma$ level of significance.


\subsection{1ES 0229+200}
\label{sec:1ES0229}

\begin{figure}[!t]
	\begin{center}

		\includegraphics[width=\columnwidth]{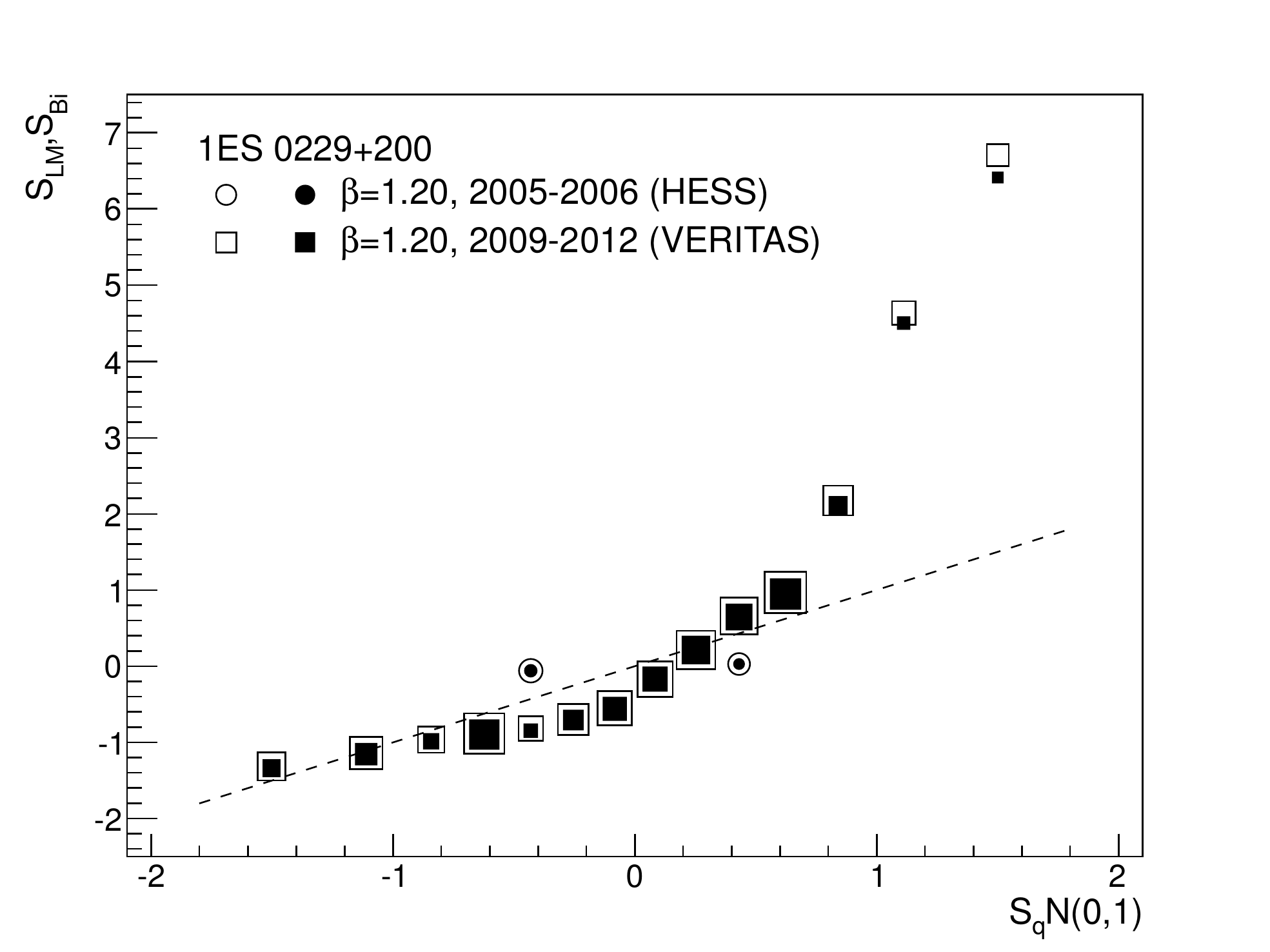}
		\caption{
			QQ--plots of sample significances for the $\gamma$--ray activity of 1ES 0229+200 are depicted.
			Circles represent the observations of the HESS~collaboration in the 2005--2006 campaign~\citep{Aharonian:2007}.
			Squares denote the VERITAS data from measurements with MJD 55118--55951~\citep{Aliu:2014}.
			Source parameter $\beta = 1.20$ was assumed.
			See also caption to Fig.~\ref{fig:PKS2005_QQ}.
		}
		\label{fig:1ES0229_QQ}
	\end{center}
\end{figure}

\par 
HBL object \es~ ($z = 0.136$) was observed by the HESS~instrument during 2005--2006~\citep{Aharonian:2007}.
No significant variations of the integral flux above the energy 580~GeV were found on any time scales.
The blazar was also a target of 2009--2012 observations of the VERITAS telescopes~\citep{Aliu:2014}.
The average integral flux above 300 GeV of about $1.7\%$ of the Crab Nebula flux was found to approach the value measured by the HESS collaboration.
An excess of the $\gamma$--ray flux was observed during October 2009 when it was almost twice as high as its average value~\citep{Aliu:2014}.
The probability of the flux being constant on annual time scales was evaluated to be $1.6\%$.
Taking into account the variations of the source activity in the X-ray domain, the VERITAS team concluded that there are indications of the flux variability in the VHE regime~\citep{Aliu:2014}.
\par
We used two annual sets of HESS data from Table~1 in Ref.~\citep{Aharonian:2007} yielding numbers of detected events in the range from over two hundred to several thousand.
We also utilized 14 VERITAS measurements from Table~1 in Ref.~\citep{Aliu:2014} collected during observing periods governed by the lunar cycle.
These observations resulted in numbers of on-- and off--source counts ranging from ten to thousands.
In QQ--plots in Fig.~\ref{fig:1ES0229_QQ}, we compare distributions of sample significances evaluated using data collected by both collaborations.
The average source paramater $\beta_{\mathrm{HESS}} = \langle \on / \alpha \off \rangle = 1.20$ used as a reference value of the blazar intensity was derived from the HESS data set.
Note that the average source parameter extracted from the VERITAS measurements is $\beta_{\mathrm{VER}} = 1.22$.
\par 
Sample statistics $\Bi$ and $\LM$ for the HESS~data (circles) do not exhibit any significant deviations from the reference diagonal.
The majority of the VERITAS~measurements of 1ES~0229+200 (squares) provide sample significances not inconsistent with the chosen source intensity.
However, two pairs of on-- and off--source counts coming from the VERITAS observations during October and November 2009 (MJD 55118--55131, 55144--55159) yield large test significances, pointing to the increase of overall source intensity.
These two instances show that the \es~ activity cannot be considered steady at least on time scales of months.
\par
\begin{figure}[!t]
	\begin{center}
		\includegraphics[width=\columnwidth]{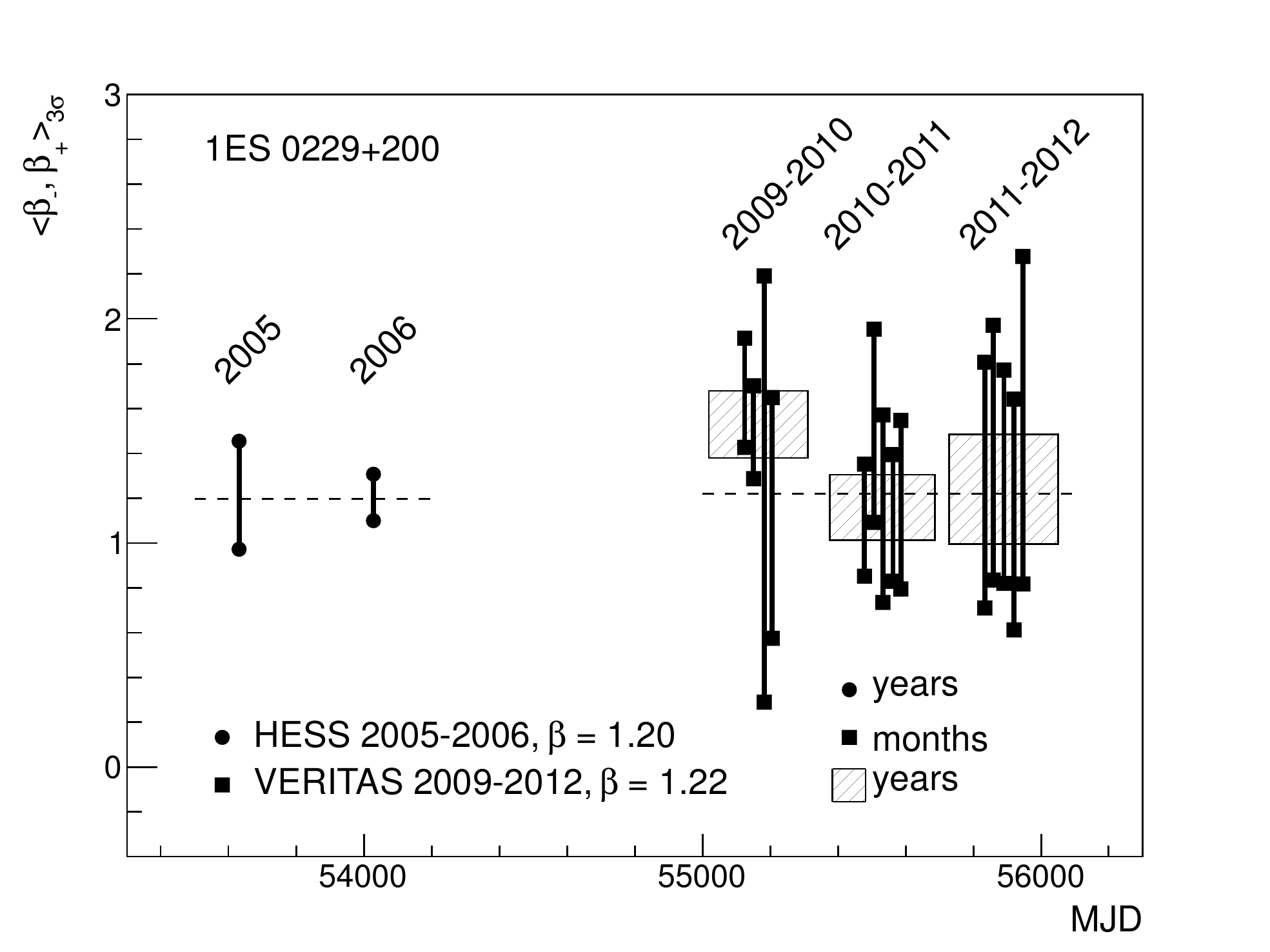}
		\caption{
			99.7\% confidence intervals for the source parameter $\beta$ of 1ES~0229+200 are depicted.
			Yearly HESS~observations~\citep{Aharonian:2007} (circles) are compared with the monthly measurements of the VERITAS collaboration~\citep{Aliu:2014} (squares).
			Hatched bands represent the confidence intervals obtained from joint sets of VERITAS data obtained in three yearly observational periods.
		}
		\label{fig:1ES0229_CI}
	\end{center}
\end{figure}

Confidence intervals for the source parameter $\beta$ at a $3\sigma$ level of significance are displayed in Fig.~\ref{fig:1ES0229_CI}.
The consistency of the source intensities measured by the HESS~(segments with circles) and VERITAS (segments with squares) collaborations, as visualized in QQ--plots in Fig.~\ref{fig:1ES0229_QQ}, is also well visible in the time sequence of confidence intervals.

\par
Monthly confidence intervals contain the average values of the source parameter derived using the HESS and VERITAS data with the only exceptions being two months in the observational period 2009--2010, see Fig.~\ref{fig:1ES0229_CI}.
Moreover, the confidence interval corresponding to October 2009 does not overlap with intervals at two latter occasions with MJD 55476--55482 and 55555--55570 observed during the 2010--2011 VERITAS campaign.
Thus, we state that the blazar varied its $\gamma$--ray emission on monthly time scales during the VERITAS observations.
The joint sets of the on-- and off--source counts taken over yearly time intervals exhibit significant excesses in the beginning of the 2009--2010 epoch (MJD 55118--55212) with respect to the HESS and VERITAS measurements of 2006 and 2010--2011.
The rest of the annual observations agree with the average source intensities deduced from both the HESS~and VERITAS data sets.
It is worth noting that the VERITAS collaboration reported the evidence of source variability supported by the 2009--2012 runs~\citep{Aliu:2014}.


\section{Conclusions}
Temporal changes in the observed $\gamma$--ray intensities of selected AGN were studied by the means of the modified on--off method with the emphasis put on its usefulness in the analyses of data gathered by Cherenkov telescopes.
The assumption of the constant source activity was ruled out consistently with the previous findings on time variability of \pks~\citep{Acero:2010} and \es~\citep{Aliu:2014}.
In particular, our results on \es~ indicate intensity changes in the 2009--2010 period, thus rejecting the long--held conjecture of its steadiness, in agreement with the conclusions of the VERITAS collaboration.
Interestingly, temporal evolution of the \pks~activity during successive months in 2006 emerges and time variability of this blazar is stated.
\par
The modified on--off scheme is not only backed by a compelling statistical motivation, but also relatively simple to implement, yet sufficiently general.
Freedom of the method from more complex calculations of fluxes makes it suitable for the examination and comparison of observed intensity changes in VHE $\gamma$--astronomy.


\section*{Acknowledgements}
This work was supported by the Czech Science Foundation grant 14-17501S.



\end{document}